\documentstyle[12pt]{article}

\begin{document}

\title{SPHINX~TT \\ Monte Carlo Program for Nucleon-Nucleon Collisions
with Transverse Polarization}
\author{OLIVER MARTIN and ANDREAS SCH\"AFER\\
	Institut f\"ur Theoretische Physik,
	J.-W. Goethe Universit\"at,\\
	60054 Frankfurt am Main, Germany}
\maketitle
\begin{abstract}
We present the program summary and long~write-up for {\sc Sphinx~tt}, a new
Monte~Carlo simulation
for transversely polarized nucleon-nucleon scattering. The program is
based on {\sc Pythia/Jetset}.
\end{abstract}

\thispagestyle{empty}

\pagebreak

{\Large\bf PROGRAM SUMMARY}

{\em Title of program:} {\sc Sphinx~tt}

{\em Catalogue number:}

{\em Program available from:} Oliver Martin,\\
${\rm http://th.physik.uni-frankfurt.de/ \sim\, martin/sphinx.html}$

{\em Computer:} IBM RS6000, PentiumPro~200 based PC; others with a
{\tt FORTRAN~77} compiler

{\em Operating system:} tested under AIX and LINUX but does not depend on
the particular operating system

{\em Programming language used:} {\tt FORTRAN~77}

{\em High speed storage required:} 5~Mbytes

{\em Number of bits per byte:} 8

{\em Number of lines in combined test deck:} 19980

{\em CPC subprograms used:} {\sc Jetset~7.3}


{\em Nature of physical problem:} This program can be used to simulate
transversely polarized nucleon-nucleon collisions at high energies.
Spins of colliding particles are taken into account. The program
allows to calculate cross sections for various processes.

{\em Method of solution:} The existing Monte Carlo program {\sc Pythia~5.6}
\cite{sjo}
has been modified to incorporate spin effects. The program incorporates
nearly all features of {\sc Pythia}.

{\em Restriction on the complexity of the problem:} The spins of the colliding
hadrons must be transverse with respect to the collision axis. (A program for
collisions of longitudinally polarized hadrons ({\sc Sphinx})
is also available.) Furthermore, the spins of the two hadrons must
be either parallel or antiparallel with respect to each other.

{\em Typical running time:} $\approx$ 0.01~ sec CPU per event on
PentiumPro~200 based PC, depends strongly on kinematical cuts

{\em References:} For general information see the {\sc Pythia} manual
\cite{sjo2}. Specific details on {\sc Sphinx~TT} (and {\sc Sphinx})
can be obtained from:
Oliver Martin, Institut f\"ur Theoretische Physik, Goethe Universit\"at,
Robert-Mayer-Str. 8-10, 60054 Frankfurt am Main, Germany,
e-mail: {\tt martin@th.physik.uni-frankfurt.de} .

\pagebreak

{\Large\bf Long Write-Up}

\section{Introduction}
During recent years the topic of transverse spin effects
has become increasingly popular. Especially the so called
transversity parton distribution $\delta q(x,Q^2)$ which is
comparable to $\Delta q(x,Q^2)$ in the case of a
longitudinally polarized nucleon
has attracted much attention (for a concise introduction see \cite{jaf}).
Due to experimental limitations,
$\delta q(x,Q^2)$ could not be measured so far. Only high energy
collions of two transversely polarized nucleons or
semi-inclusive deep inelastic electron-nucleon scattering with transverse
polarization allow a determination of $\delta q(x,Q^2)$.

However, in 1992 the RHIC Spin Collaboration was constituted \cite{bnl}. One
of its major projects is the measurement of the transversity parton
distribution by analysing collisions of transversely polarized protons
at RHIC, which is now under construction at BNL.
RHIC will be able to scatter protons at cm energies of up to
$\sqrt{s}=500\,{\rm GeV}$ with high transverse or longitudinal polarization
of up to 70\% . A first measurement may already
be possible in the year 2000 or shortly after. To plan for such
experiments requires extensive Monte~Carlo simulations. {\sc
Sphinx~TT} was developped to perform such studies.

Developing a state-of-the-art Monte Carlo simulation takes
several years. Therefore, we
used {\sc Sphinx} \cite{guelle} which itself
is an extension of the popular {\sc Pythia~5.6/Jetset~7.3}
\cite{sjo} as a basis for our program.
{\sc Sphinx} is able to simulate collisions of {\em two
longitudinally} polarized
nucleons in the framework of the QCD improved parton model and
describes nucleons on twist-2 level \cite{jaf}.
On this level no new physics is involved in collisions of
a transversely polarized nucleon with a longitudinally polarized
nucleon \cite{ral}. For this reason we decided only to cover
the scattering of {\em two transversely} polarized protons with our
program, which we consequently called {\sc Sphinx~tt}.
{\sc Sphinx} is an acronym for {\em Simulator of Polarized Hadronic
INteract(X)ions}.

The features of {\sc Sphinx~tt} are similar to those of {\sc Sphinx},
namely
\begin{itemize}

\item Spin propagation. The spin information is propagated until the
hard partonic interaction, i.e. the spin of the partons is taken into account
in the initial state showering and hard partonic scattering.
Final state interactions, such as final
state showering and fragmentation, are treated as being spin-independent.

\item Polarized hard processes are described at leading order in the
electromagnetic and strong coupling constants.
The processes currently implemented in the
polarized mode are summarized in Table~\ref{tab1}. Contact interactions
are not available yet in the polarized mode.

\item
Polarized parton distributions. Two model parametrizations for
$\delta q(x,Q^2)$ are provided, the inclusion of new ones is easy.

\item
Documentation. Spin information has been added to the event listing.

\item {\sc Sphinx~tt} is equivalent to {\sc Pythia~5.6} in its
unpolarized mode.
\end{itemize}

Hereafter we do not try to explain the structure of {\sc Pythia 5.6}
or the concepts that lead to the creation of {\sc Sphinx}.
Readers which are
unfamiliar with these programs should consult \cite{sjo2,guelle}.
In the next section we will present the most important differences between
{\sc Sphinx} and {\sc Sphinx tt}. Section~\ref{sub} contains a detailed
description of the modified subroutines as well as three tables
with all new switches and internal variables. Finally an example
main program is presented.

\begin{table}
\begin{center}
\begin{tabular}{|r|l|l|}
\hline
ISUB & Process & Comment \\
\hline\hline
11   & $q_i q_j \rightarrow q_i q_j$ &
(anti-)quark -- (anti-)quark scattering;\\
     & & annihilation is not included\\
\hline
12   & $q_i \bar q_i \rightarrow q_k \bar q_k$ & annihilation process \\
\hline
13   & $q_i \bar q_i \rightarrow gg$ &
annihilation into gluon pair \\
\hline
14   & $q_i \bar q_i \rightarrow g\gamma$&
annihilation into gluon and prompt $\gamma$\\
\hline
18   & $q_i \bar q_i \rightarrow \gamma \gamma$ &
annihilation into $\gamma$-pair\\
\hline
28   & $q_i g \rightarrow q_i g$&
(anti-)quark -- gluon scattering\\
\hline
29   & $q_i g \rightarrow q_i \gamma$ &
prompt $\gamma$-production in (anti-)quark -- \\
     &  & gluon scattering\\
\hline
53   & $gg \rightarrow q_k \bar q_k$ & gluon fusion \\
\hline
68   & $gg \rightarrow gg$ & gluon -- gluon scattering\\
\hline
\rule[-0mm]{0mm}{5mm}
191  & $q_i \bar q_i \rightarrow f_k \bar f_k$& annihilation into
lepton-pair\\
& & or quark -- (anti-)quark pair \\
& & (Drell-Yan process); this process is new\\
& &  and equivalent to the $\gamma$-piece of\\
& & ISUB=1 in {\sc Pythia} \\
\hline
\end{tabular}
\caption{\label{tab1}List of processes implemented in the polarized mode}
\end{center}
\end{table}


\section{\label{sub}From Sphinx to Sphinx~TT: modifications}
The most notable difference between the scattering of two {\em longitudinally}
polarized nucleons and the scattering of two {\em transversely} polarized ones
is that the symmetries in the cm frame are different. In the former case the
system is invariant under rotations around the beam axis whereas in the
latter case the transverse spin directions and the beam axis specify
one or two planes in space. As a result, the rotational invariance is broken
and the partonic cross sections now also may depend on the azimuthal
scattering
angle.

In {\sc Sphinx~tt} the spins of the colliding nucleons are required to be
either parallel or anti-parallel. Allowing for arbitrary transverse spins
only changes the azimuthal dependence of the cross sections in a
trivial way \cite{ral}.

Since the partonic cross sections in {\sc Pythia} and {\sc Sphinx} are
independent of the azimuth $\phi$ it simply can be chosen according to a
flat distribution. In comparison, the generation of the remaining
kinematical variables is a complicated process.
In order to make as few modifications as possible the
event machinery in {\sc Sphinx~tt} was not rewritten but a trick was used
to include the correct generation of the azimuth.

During the initialization {\sc Pythia/Sphinx} searches
for the absolute maxima of the various selected partonic subprocesses. If
``by accident'' the program finds a point in the kinematical space
with a larger partonic cross section, the program adjusts the maximum value.
For that reason, one can simply include the normal $\phi$-dependent
cross sections and all kinematical variables will be chosen correctly.

This method only works for $2\rightarrow 2$-processes. Unfortunately,
in {\sc Pythia} / {\sc Sphinx} the Drell-Yan process
$q_i \bar q_i \rightarrow l  \bar l$ was implemented as a
$2\rightarrow 1$-process with a successive decay of the gauge boson into
a lepton pair. Furthermore the two steps were performed by different
subroutines. It was impossible to implement transverse polarization here
without violating the program structure strongly and  we simply included
a new $2\rightarrow 2$ subprocess for Drell-Yan events.
For similar reasons the subprocesses which cover the QCD corrections to
the simple Drell-Yan graph could not be made available.

The modification of the initial state showering routine was relatively simple.
Only the appropriate splitting functions had to be included and the
handling of the Lorentz boosts required some changes in order to
treat the azimuth correctly.

\begin{table}
\begin{center}
\begin{tabular}{|l|rl|c|}
\hline
Parameter & & Description & Default \\
\hline
\hline
{\tt MSTP(171)} &     & beam polarization & 0 \\
                & =0:  & unpolarized & \\
		& =1:  & polarization in $+x$ direction &\\
		& =2:  & polarization in $-x$ direction &\\
\hline
{\tt MSTP(172)} &     & target polarization & 0 \\
		& =0:  & unpolarized & \\
		& =1:  & polarization in $+x$ direction & \\
		& =2:  & polarization in $-x$ direction& \\
\hline
{\tt MSTP(175)} &     & use of polarized parton distributions in & 1\\
		&     &	polarized initial state shower ({\tt MSTP(176)=1})& \\
		& =0:  & unpolarized distribution; for testing only;& \\
	 	&     & {\em do not use!}         &  \\
		& =1:  & polarized & \\
\hline
{\tt MSTP(177)}&& set of polarized parton distributions $\delta q(x,Q^2)$&0\\
		&     &  used; in addition, one has to specify& \\
 		&     &  an unpolarized set as in standard {\sc Pythia} & \\
		& =0: & $\delta q(x,Q^2)=0$ (no polarization) & \\
		& =1: & maximal polarization; $\delta q(x,Q^2)=q(x,Q^2)$ & \\
		& =2: & Model A: $\delta q(x,4\, {\rm GeV}^2) =
			\Delta q(x,4\,{\rm GeV}^2$)              & \\
		&     & + correct DGLAP evolution      & \\
		&     & set by Bartelski-Tatur \cite{bartat} is used	& \\
		& =3: & same as before but Gehrmann-Stirling & \\
                &     & lo set C \cite{gehsti} is used      & \\
\hline
{\tt MSTP(178)} && percentage of beam and target polarization & 0 \\
\hline
{\tt MSTP(179)} && switch off polarization temporarily in {\tt PYSIGH} & \\
		&& and {\tt PYSTFU} resp. & \\
		&=0:& no action & \\
		&=1:& switch off polarization &\\
\hline
{\tt MSTP(180)} && mode selection (master switch) 	& 0 \\
		& =0: & unpolarized mode; this value overrides all& \\
		&     & other polarization switches	& \\
		& =1: & polarized mode	& \\
\hline
{\tt NSUB(ISUB)} && mode for subprocess {\tt ISUB}	& 0 \\
		& =0: & unpolarized treatment		& \\
		& =1: & polarized treatment		& \\
\hline
{\tt NSEL}	& & menu of polarized processes	& 0 \\
		& =1: & {\tt ISUB = 11,12,13,28,53,68} switched on & \\
		& =10: & {\tt ISUB = 14,18,29} switched on & \\
\hline

\end{tabular}
\caption{\label{tab2}Parameters controlling the polarized mode}
\end{center}
\end{table}

\begin{table}
\begin{center}
\begin{tabular}{|l|rl|c|}
\hline
Variable & & Description & Com. Block \\
\hline
\hline
{\tt MINT(311)} & & beam polarization & {\tt PYINT1} \\
		& =0: & unpolarized & \\
		& =1: & transverse polarization in $+x$ direction & \\
		& =2: & transverse polarization in $-x$ direction & \\
\hline
{\tt MINT(312)} & & target polarization & {\tt PYINT1} \\
		& =0: & unpolarized & \\
		& =1: & transverse polarization in $+x$ direction & \\
		& =2: & transverse polarization in $-x$ direction & \\
\hline
{\tt MINT(313)} & & polarization of shower initiator on beam side&
{\tt PYINT1} \\
		& =0: &unpolarized & \\
		& =1: & transverse polarization in $+x$ direction & \\
		& =2: & transverse polarization in $-x$ direction & \\
\hline
{\tt MINT(314)} & & polarization of shower initiator on target side &
{\tt PYINT1} \\
		& =0: & unpolarized & \\
		& =1: & transverse polarization in $+x$ direction & \\
		& =2: & transverse polarization in $-x$ direction & \\
\hline
{\tt MINT(315)} & & polarization of hard interacting parton &{\tt PYINT1} \\
		& & on beam side & \\
		& =0: &unpolarized & \\
		& =1: & transverse polarization in $+x$ direction & \\
		& =2: & transverse polarization in $-x$ direction & \\
\hline
{\tt MINT(316)} & & polarization of hard interacting parton & {\tt PYINT1} \\
		& & on target side & \\
		& =0: & unpolarized & \\
		& =1: & transverse polarization in $+x$ direction & \\
		& =2: & transverse polarization in $-x$ direction & \\
\hline
{\tt VINT(397)} & & azimuthal angle of the hard scattering plane&
{\tt PYINT1} \\
                & & with respect to the x-axis before the initial state& \\
		& & shower and primordial transverse momentum & \\
		& & are generated; measured in cm-frame of hard& \\
		& & interaction & \\
\hline
{\tt VINT(398)} & & same as {\tt VINT(397)} after generation of initial &
{\tt PYINT1}   \\
		& & state shower and prim. transverse momentum & \\
\hline
{\tt VINT(399)} & & scattering angle in the cm frame of the hard &
{\tt PYINT1} \\
		& & partonic interaction before the initial state& \\
		& & shower and primordial transverse momentum & \\
		& & are generated& \\
\hline
{\tt VINT(400)} & & same as {\tt VINT(399)} after generation of initial &
{\tt PYINT1}   \\
		& & state shower and prim. transverse momentum & \\
\hline
\end{tabular}
\caption{\label{tab3}Internal variables storing polarization information}
\end{center}
\end{table}

\begin{table}
\begin{center}
\begin{tabular}{|l|rl|c|}
\hline
Variable & & Description & Com. Block \\
\hline
\hline
{\tt ISIG(1000,6)} & & hard scattering information of {\tt N}th channel &
{\tt PYINT3} \\
{\tt ISIG(N,1)} & & particle code of {\tt N}th channel on beam side & \\
{\tt ISIG(N,2)} & & particle code of {\tt N}th channel on target side & \\
{\tt ISIG(N,3)} & & colour flow of {\tt N}th channel& \\
{\tt ISIG(N,4)} & & polarization of {\tt N}th channel on beam side & \\
{\tt ISIG(N,5)} & & polarization of {\tt N}th channel on target side & \\
{\tt ISIG(N,6)} & & not used & \\
\hline
{\tt KD(I)}	 & & polarization of {\tt I}th line in the & {\tt DPYPOL} \\
                 & & event listing & \\

		 & =0: & no polarization	& \\
		 & =1: & transverse polarization in $+x$ direction & \\
		 & =2: & transverse polarization in $-x$ direction & \\
\hline
{\tt XSFX(2:-40:40,0:2)} & & $x$ times parton distribution for given &
{\tt PYINT3} \\
		& & $x$ and $Q^2$ of flavour {\tt KFL= -40:40} & \\
		& & and transversity {\tt KFLD=0:2} on beam side& \\
		& & ({\tt JT=1}) and target side ({\tt JT=2}) resp. & \\
{\tt XSFX(JT,KFL,0)}&& unpolarized & \\
{\tt XSFX(JT,KFL,1)}&& positive polarization with respect to & \\
&&transverse hadron polarization&\\
{\tt XSFX(JT,KFL,2)}&& negative polarization with respect to & \\
&&transverse hadron polarization&\\
\hline
\end{tabular}
\caption{Internal variables storing polarization information}
\end{center}
\end{table}


\section{Common Blocks and Subroutines}
In this section a detailed description of the modified subroutines is given.
In order for this section to be self contained we present the differences
between {\sc Pythia} and {\sc Sphinx~tt} rather than giving a
summary of the differences between {\sc Sphinx} and {\sc Sphinx tt}.
The general structure of the code as well as the unchanged
parameters and variables are not explained.
However, it is shown in detail how the local polarization switch
{\tt IPOL} is implemented into the different subroutines.
Only the polarized case
({\tt IPOL=1}) will be discussed, because in the unpolarized case
({\tt IPOL=0}) each subroutine works exactly as the corresponding
{\sc Pythia~5.6} subroutine. The subroutines of
{\sc Sphinx~tt} which we don't mention explicitly remain
unchanged with respect to {\sc Pythia~5.6}.

To incorporate polarization the following common blocks have either
been enlarged,
replacing the corresponding {\sc Pythia} common blocks or have been added:

\begin{itemize}
\item {\tt COMMON/PYINT3/XSFX(2,-40:40,0:2),ISIG(1000,6),SIGH(1000)}
\item {\tt COMMON/PYSUBS/MSEL,NSEL,MSUB(200),NSUB(200),KFIN(2,-40:40),
CKIN(200)}
\item {\tt COMMON/DYPOL/KD(4000)}
\end{itemize}

Information about the new internal variables and enlarged arrays can be found
in Table \ref{tab2}.

{\tt MAIN PROGRAM}
\newline
{\bf Purpose:} to set up the polarized event generation. The variables which
have to be specified are listet in Table~\ref{tab2}.
\newline
{\bf Remarks:} An example program is given in Section~\ref{demo}

{\tt SUBROUTINE PYINIT}
\parskip0cm
\newline
{\bf Purpose:}
\begin{itemize}
\item to display {\sc Sphinx~tt} header;
\item to check partially the
availability of the desired polarization scenario, i.e. to check that the
master switch for polarization {\tt MSTP(180)} is set properly, that the
selected partonic subprocesses can be treated polarized and to control and
compose the polarization menue via {\tt NSEL};
\item to call {\tt DPLIST}
instead of {\tt LULIST} (see below).
\end{itemize}
{\bf New Parameters:} {\tt MSTP(180), NSEL, NSUB(ISUB)}
\newline
{\bf Internal Polarization Switch:} {\tt IPOL=MSTP(180)}
\newline
{\bf Remarks:} If the chosen scenario is not allowed, the program
is terminated with an appropriate error message.
\parskip4.0ex plus 0.5ex minus 0.5ex

{\tt SUBROUTINE PYEVNT}
\parskip0cm
\newline
{\bf Purpose:}
\begin{itemize}
\item to start polarized event generation;
\item  to call
{\tt DPEDIT} instead of {\tt LUEDIT} (see below).
\end{itemize}
{\bf New Parameters:} {\tt MSTP(180)}
\newline
{\bf New Variables:} {\tt VINT(397), VINT(398),
VINT(399), VINT(400)}
\newline
{\bf Internal Polarization Switch:} {\tt IPOL=MSTP(180)}
\newline
{\bf Remarks:} {\tt VINT(397)-VINT(400)} can be used to measure
the rotation of the hard scattering partonic system caused by the
initial state shower and the primordial transverse momentum of
the partons; if this effect is not small {\em (check!)}
 the transversity distribution
is not used correctly in the hard scattering because the particle spins
are no longer aligned along the $x$-axis; also, the implementation of
the initial state shower is then  no longer appropriate since at each
partonic branching the azimuthal angles are chosen according to a
flat distribution instead of the correct one. This latter
shortcoming could be overcome if needed, but this would require major
changes in the program.
\parskip4.0ex plus 0.5ex minus 0.5ex

{\tt SUBROUTINE PYINKI(CHFRAM,CHBEAM,CHTARG,WIN)}
\parskip0cm
\newline
{\bf Purpose:}
\begin{itemize}
\item to check availability of the desired hadronic polarization
scenario, i.e. to control that the selected hadron can be treated polarized;
\item to store the polarization of beam and target for the event listing in
{\tt KD(1)} and {\tt KD(2)} and for internal use in {\tt MINT(311)} and
{\tt MINT(312)}.
\end{itemize}
{\bf New Parameters:} {\tt KD(I), MSTP(171), MSTP(172), MSTP(180)}
\newline
{\bf New Internal Variables:} {\tt MINT(311), MINT(312)}
\newline
{\bf Internal Polarization Switch:} {\tt IPOL=MSTP(180)}
\newline
{\bf Remarks:} At present only nucleons and their antiparticles can be treated
polarized. If a not allowed
scenario has been chosen, the program is terminated
with an appropriate error message.
\parskip4.0ex plus 0.5ex minus 0.5ex

{\tt SUBROUTINE PYRAND}
\parskip0cm
\newline
{\bf Purpose:}
\begin{itemize}
\item to adapt {\tt PYRAND} to the new enviroment - all relevant
arrays which have been enlarged or added to the common blocks in other
subroutines are modified here as well;
\item to extend event shape selection to
incorporate transverse spin;
\item to store polarization of the partons
entering the hard interaction in {\tt MINT(313) - MINT(316)}
(see Table~\ref{tab3}).
\end{itemize}
{\bf New Parameters:} {\tt MSTP(180), NSUB(ISUB)}
\newline
{\bf New internal variables:} {\tt MINT(313),MINT(314),MINT(315),MINT(316)}
\newline
{\bf Internal polarization switch:} {\tt IPOL=MSTP(180)$\times$NSUB(ISUB)}
\newline
{\bf Remarks:} Note that {\tt MINT(313)=MINT(315)} and
{\tt MINT(314)=MINT(316)} but the values of {\tt MINT(313)} and {\tt
MINT(314)}
are changed later by the initial state shower in {\tt PYSSPA}.
\parskip4.0ex plus 0.5ex minus 0.5ex

{\tt SUBROUTINE PYSCAT}
\parskip0cm
\newline
{\bf Purpose:}
\begin{itemize}
\item to adopt {\tt PYSCAT} to the new environment (see
{\tt PYRAND});
\item to store transverse spin information of the partons entering
the hard interaction;
\item to add spin information to lines 1, 2 and 5, 6 in the event
listing (see below).
\end{itemize}
{\bf New Parameters:} {\tt KD(I), MSTP(180), NSUB(ISUB)}
\newline
{\bf New internal variables:} {\tt MINT(315), MINT(316)}
\newline
{\bf Internal Polarization Switch:} {\tt IPOL=MSTP(180)$\times$NSUB(ISUB)}
\parskip4.0ex plus 0.5ex minus 0.5ex

{\tt SUBROUTINE PYSSPA(IPU1,IPU2)}
\newline
\parskip0cm
{\bf Purpose:}
\begin{itemize}
\item to perform polarized initial state showering, transversity
dependent DGLAP evolution equations are used in the backward evolution
algorithm;
\item to enlarge all relevant arrays in an appropriate manner to
incorporate polarization;
\item to check proper selection of the polarized
initial state shower scenario, i.e. to control that
{\tt MSTP(175)} and {\tt MSTP(176)}
are set correctly;
\item to store the polarization of the initial state shower
initiators ({\tt MINT(313), MINT(314)}).
\end{itemize}
{\bf New Parameters:} {\tt KD(I), MSTP(171), MSTP(172), MSTP(175),
\newline MSTP(176),
MSTP(180), NSUB(ISUB)}
\newline
{\bf New Internal Variables:} {\tt MINT(313),MINT(314), MINT(315), MINT(316)}
\newline
{\bf Internal Polarization Switch:} {\tt IPOL=MSTP(180)$\times$NSUB(ISUB)}
\newline
{\bf Remarks:}
At the present stage only QCD shower can be treated polarized, QED showering
has to be done in the unpolarized manner. The combination
{\tt MSTP(175)=0} and {\tt MSTP(176)=1} allows to simulate
{\em polarized} showering with the use of {\em unpolarized} parton
distributions. This option is just for testing and should not be selected by
the user! If {\tt MSTP(175)} or {\tt MSTP(176)} are set improperly,
the program stops with an appropriate error message. The internal variables
{\tt MINT(313)} and {\tt MINT(314)} are changed to their final values
in this subroutine. \newline
The handling of the Lorentz boosts in {\tt PYSSPA} has been
modified so that the azimuthal angle is propagated through the event.
Nevertheless, at
each partonic branching the azimuth of the decay plane with respect to the
transverse spin of the decaying parton is distributed according to a
flat distribution instead of the correct one.
Only if the initial state shower
does not cause a substantial rotation of the hard scattering partonic
system this approximation is justified. This shortcoming can be overcome
if needed but would require substantial changes in the program.
\parskip4.0ex plus 0.5ex minus 0.5ex

{\tt SUBROUTINE PYMULT(MMUL)}
\newline
{\bf Purpose:} to switch off polarization in {\tt PYSIGH}
(set {\tt MSTP(179)=1} temporarily) when called from {\tt
PYMULT}. This is done because multiple interactions cannot be
treated polarized at the moment.
\newline
{\bf New Parameters:} {\tt MSTP(179)}

{\tt SUBROUTINE PYREMN(IPU1,IPU2)}
\newline
\parskip0cm
{\bf Purpose:}
\begin{itemize}
\item to adopt {\tt PYREMN} to the new environment
(see {\tt PYRAND});
\item to add spin information to  lines 3 and 4 in the event listing
(see below).
\end{itemize}
{\bf New Parameters:} {\tt KD(I)}
\newline
{\bf Remarks:} The generation of the primordial transverse momentum has been
modified so that the azimuthal angles are now treated correctly.
\parskip4.0ex plus 0.5ex minus 0.5ex

{\tt SUBROUTINE PYSIGH(NCHN,SIGS,IOC)}
\parskip0cm
\newline
{\bf Purpose:}
\begin{itemize}
\item  to evaluate the transversity dependent hadronic cross sections
by convolution of the spin dependent parton distributions with the
transversity dependent partonic cross sections;
\item to supply the
subroutine with the transversity dependent partonic cross sections.
\end{itemize}
{\bf New Parameters:} {\tt IOC,MSTP(171),MSTP(172),MSTP(179),MSTP(180),
\newline
NSUB(ISUB)}
\newline
{\bf Internal Polarization Switch:} \newline
{\tt IPOL=MSTP(180)$\times$NSUB(ISUB)$\times$(1-MSTP(179))}
\newline
{\bf New Partonic Subprocess:} {\tt ISUB=191} ($q_i +\bar q_i
\rightarrow f_k \bar f_k$)
\newline
{\bf Remarks:}
{\tt PYSIGH} will always run in the unpolarized mode when it is called by
{\tt PYMULT} which sets temporarily {\tt MSTP(179)=1} in {\tt PYSIGH}.
When one evaluates the spin dependent hadronic cross sections
one has to be aware that the hadrons are labelled according to their
absolute spin whereas the partons are specified
according to their relative spin with respect to the spin of the hadron.
The parton distributions are passed  from {\tt PYSTFU} to {\tt PYSIGH}
through the array {\tt XPQ(KFL,KFLD)} (see below) and stored in the
array {\tt XSFX(N,KFL,KFLD)}, where {\tt KFLD} denotes the transversity.
{\tt ISIG(N,I)} contains the information about the {\tt N}th reaction
channel of the chosen partonic subprocess.
The new entries {\tt ISIG(N,4)} and {\tt ISIG(N,5)}
specify the transverse spin of the partons at the beam and target side
respectively. {\tt ISIG(N,6)} is reserved but not used at the moment.
\newline
The cross sections in {\sc Pythia} and {\sc Sphinx} are independent
of the azimuthal scattering angle which therefore can be easily chosen
according to a flat distribution. Contrarily, in {\sc Sphinx~tt} the
cross section depends on the azimuth $\phi$.
Correct generation can be achieved by simply using the $\phi$-dependent
formulae of the partonic cross sections. Furthermore, during the
initial search for the maximum of the partonic cross sections the
azimuth must be fixed ({\tt IOC=0}). During the normal event generation
{\tt IOC=1} is set by {\tt PYRAND}.
No other changes are required to ensure  correct generation of all
kinematical variables.
Contact interactions are not yet available for the polarized mode.
 The new partonic subprocess {\tt ISUB=191} that corresponds to {\tt ISUB=1}
had to be
included in {\tt PYSIGH} in order to preserve the program structure.
The important difference is that
no $Z^0$-propagator is included in the new process and that
{\tt ISUB=191} is a $2\rightarrow 2$-process rather than a $2\rightarrow 1$
one.
\newline
One should be careful when using the subprocess {\tt ISUB=11} because
here only interference graphs carry an asymmetry. These are only
taken into account if one sets {\tt MSTP(34)=1}.
\parskip4.0ex plus 0.5ex minus 0.5ex

{\tt SUBROUTINE PYSTFU(KF,X,Q2,XPQ)}
\newline
{\bf Purpose:} to evaluate the transversity dependent parton distributions
for given flavour ({\tt KF}), $x$ ({\tt X}), and $Q^2$ ({\tt Q2})
according to the
selected parametrizations and models
\newline
{\bf New Parameters:} {\tt MSTP(177), MSTP(178), MSTP(179), MSTP(180),
\newline MSUB(ISUB)}
\newline
{\bf Internal Polarization Switch:}
{\tt IPOL=MSTP(180)$\times$(1-MSTP(179))}
\newline
{\bf Remarks:}
Call of {\tt PYSTFU} returns $x$ times the parton distribution
functions for given flavour, $x$, and $Q^2$ for both spin orientations and
an averaged (unpolarized) value. The values are stored in the
array {\tt XPQ(KFL,KFLD)} which has been enlarged from {\tt XPQ(-25:25)}
to {\tt XPQ(-25:25,0:2)}. {\tt XPQ(KFL,0)} contains the unpolarized
distributions,  {\tt XPQ(KFL,1)} distributions for
positive relative transversity and {\tt XPQ(KFL,2)} distributions for
negative relative transversity. The parton distributions are
selected by switches described earlier in Table~\ref{tab2}
The polarized distributions $q_{\uparrow\downarrow}(x,Q^2)$ are constructed
from the unpolarized  ones $q(x,Q^2)$
and the transversity distributions
$\delta q(x,Q^2)$ selected by {\tt MSTP(177)}
according to $q_{\uparrow\downarrow}=\frac{1}{2}(q\pm \delta q)$.
The interface to the
CERN parton distribution library has been updated so that
{\sc Pdflib 4.0} and higher versions can be used with {\sc Sphinx~tt}.
Two subroutines have been added to calculate the polarized
distributions, namely {\tt BARTAT} and {\tt GEHRMA}. They require the
data files {\tt bartat.dat} and {\tt gehrmann.dat} which are supplied with the
program and must be
visible to the {\tt FORTRAN open} statement. Therefore the paths of these
files have to be modified appropriatly in the subroutines mentioned above.
When $q$ and $\delta q$ are combined to compute the distributions for
fixed transversity an unitarity check is performed --- if one of the
resulting values is negative this value is set to zero and the value
for the other spin direction is set equal to the unpolarized one.
Only polarized proton
parametrizations are implemented. Neutron distributions
are obtained by SU(2) symmetry. Charge conjugation is used to describe
the corresponding antiparticles.

{\tt SUBROUTINE PYSTPR(X,Q2,XPPR)}
\newline
{\bf Purpose:} to calculate the unpolarized parton distributions of the
proton;
to update the implementation of the {\sc Cteq2} parton distribution functions
\newline
{\bf Remark:} the Interface to {\tt PYCTQ2} was taken from {\sc Pythia~5.7}

{\tt FUNCTION PYCTQ2 (Iset, Iprt, X, Q)}
\newline
{\bf Purpose:} to give the revised {\sc Cteq2} parton distribution sets
with extended range in
parametrized form.
\newline
{\bf Remark:} This function was taken from {\sc Pythia~5.7}.

{\tt SUBROUTINE BARTAT(X,Q2,UPV,DNV,EM,DELTA)}
\newline
{\bf Purpose:} to return $x$ times the polarized parton distributions
evaluated at given $x$ ({\tt X}) and $Q^2$ ({\tt Q2}) for model~1.
{\tt UPV, DNV, EM, DELTA} contain $\delta u_v$, $\delta d_v$,
$\delta M$ and $\delta \delta$ \cite{bartat} repectively.
\newline
{\bf Remarks:}
We use
$\delta q(x,4\, {\rm GeV}^2)=\Delta q(x,4\, {\rm GeV}^2)$ for lack of
any knowledge of $\delta q(x,Q^2)$. One actually has to expect that
$\delta q$ and $\Delta q$ differ substantially because their evolution
equations are different. For $\Delta q(x,4{\rm GeV}^2)$
the Bartelski-Tatur parametrization \cite{bartat}
is used.
The $\delta q$-values
for higher $Q^2$ were obtained by correct DGLAP evolution.
{\tt BARTAT} requires the data file {\tt bartat.dat}.

{\tt SUBROUTINE GEHRMA(X,Q2,UPV,DNV,EM)}
\newline
{\bf Purpose:} to return $x$ times the polarized parton distributions
evaluated at given $x$ ({\tt X}) and $Q^2$ ({\tt Q2}) for model~2.
{\tt UPV, DNV, EM} contain $\delta u_v$, $\delta d_v$ and
$\delta s$ repectively.
The sea is assumed to be SU(3) symmetric.
\newline
{\bf Remarks:}
In this model
$\delta q(x,4\, {\rm GeV}^2)=\Delta q(x,4\, {\rm GeV}^2)$ is assumed
for lack of any knowledge of $\delta q$. The Germann-Stirling Set C
\cite{gehsti} is used for $\Delta q$.
The $\delta q$-values for higher $Q^2$ were
obtained by correct DGLAP evolution.
{\tt GEHRMA} requires the data file {\tt gehrmann.dat}.

{\tt SUBROUTINE DPLIST(MLIST)}
\newline
{\bf Purpose:} to display the polarizations of the particle in the event
listing.
\newline
{\bf New Parameters:} {\tt KD(I)}
\newline
{\bf Remarks:} {\tt DPLIST} is a modification of the {\sc Jetset}
subroutine {\tt LULIST}. It is changed to display the polarization
in the final listing. The sign displayed just behind the particle code
denotes polarization with respect to the $x$-axis. When the sign is missing
the particle has been treated as being unpolarized. The information
is taken from the vector {\tt KD(I)} and printed using the symbols
\parskip0cm
\begin{description}
\item[= ` ']: no polarization ({\tt KD(I)=0})
\item[= `$\wedge$']: transverse polarization in $+x$ direction ({\tt KD(I)=1})
\item[= `v']: transverse polarization in $-x$ direction ({\tt KD(I)=2})
\end{description}
\parskip4.0ex plus 0.5ex minus 0.5ex

{\tt SUBROUTINE DPEDIT(MEDIT)}
\newline
{\bf Purpose:} to properly compress the vector {\tt KD(I)} which contains the
polarization information.
\newline
{\bf New Parameters:} {\tt KD(I)}
\newline
{\bf Remarks:} {\tt DPEDIT} is a modification of the {\sc Jetset} subroutine
{\tt LUEDIT}.

{\tt SUBROUTINE PDFSET(PARM,VALUE)}
\newline
{\bf Purpose:} new dummy routine for new {\sc Pdflib}
interface; has to be removed
if {\sc Pdflib} is used.

{\tt SUBROUTINE STRUCTM(XX,QQ,UPV,DNV,USEA,DSEA,STR,CHM,BOT,TOP,GLU)}
\newline
{\bf Purpose:} new dummy routine for new {\sc Pdflib} interface; has to be
removed
if {\sc Pdflib} is used.

\section{\label{demo}The Main Program}
\subsection{The code}
\begin{verbatim}
      PROGRAM EXAMPLE

C     Example of a Main Program for event generation
C     in proton-proton scattering with transverse polarization
C
C     One beam is polarized in +x direction, the other one in
C     -x direction
C
C     This program has to be linked with
C     the programs SPHINX TT and JETSET 7.3,
C     the data files BARTAT.DAT and GEHSTI.DAT, and
C     the CERN Libraries

C     COMMON BLOCKS of SPHINX TT for event generation
	
      COMMON/LUDAT1/MSTU(200),PARU(200),MSTJ(200),PARJ(200)
      COMMON/LUJETS/N,K(4000,5),P(4000,5),V(4000,5)
      COMMON/PYSUBS/MSEL,NSEL,MSUB(200),NSUB(200),
     &              KFIN(2,-40:40),CKIN(200)
      COMMON/PYPARS/MSTP(200),PARP(200),MSTI(200),PARI(200)
      COMMON/PYINT5/NGEN(0:200,3),XSEC(0:200,3)
      COMMON/PAWC/HBOOK(10000)

      DIMENSION ROBO(3)

C=============================================================
C     polarization set up
C=============================================================

C     polarized simulation
      MSTP(180)=1

C     beam polarized in +x direction
      MSTP(171)=1

C     "target" polarized in -x direction
      MSTP(172)=2

C     "fully" polarized parton distributions
      MSTP(177)=1

C     the degree of polarization of each beam is only 70%
      MSTP(178)=70

C     polarized initial state shower
      MSTP(176)=1

C=============================================================
C     event set-up
C=============================================================

C     select hard subprocesses a la carte
      MSEL=0

C     choose "Drell-Yan" q q~ --> gamma --> f f~
      MSUB(191)=1

C     switch on polarization for 191
      NSUB(191)=1

C     switch off multiple interactions
      MSTP(81)=0
      MSTP(131)=0

C     number of events to be generated
      NEVENT=1000

C=============================================================
C     start
C=============================================================

C     initialize
      CALL PYINIT('CMS','p','p',200.)

C     book histograms
      CALL HLIMIT(10000)
      CALL HBOOK1(1,'azimuthal distrib.',20,-3.1416,3.1416,0.)

C     event loop
      DO 200 I=1,NEVENT

        CALL PYEVNT

C       list first few events
        IF(I.LE.6) CALL DPLIST(1)

C       detect myons and electrons
        IF(ABS(K(7,2)).EQ.13.OR.ABS(K(7,2)).EQ.11) THEN

C       boost to di-lepton rest-frame
        DO 100 J=1,3
          ROBO(J)=(P(7,J)+P(8,J))/(P(7,4)+P(8,4))
100     CONTINUE
        ROBOT=ROBO(1)**2+ROBO(2)**2+ROBO(3)**2
        IF(ROBOT.GE.0.999999) THEN
          ROBOT=1.00001*SQRT(ROBOT)
          ROBO(1)=ROBO(1)/ROBOT
          ROBO(2)=ROBO(2)/ROBOT
          ROBO(3)=ROBO(3)/ROBOT
        ENDIF
        CALL LUDBRB(1,N,0.,0.,-DBLE(ROBO(1)),
     &    -DBLE(ROBO(2)),-DBLE(ROBO(3)))

C	calculate azimuthal angle
        PHI=PLU(7,15)

C       fill histogram
        CALL HFILL(1,PHI,0.,1.)
        ENDIF

200   CONTINUE

C     print cross section and statistics
      CALL PYSTAT(1)
      CALL HRPUT(0,'example.hbk','n')

      END

\end{verbatim}

\subsection{The Event Listing}
\begin{verbatim}


                         SPHINX TT  version 1.00
               **  Last date of change:  10 Oct 1996  **




               The Lund Monte Carlo - PYTHIA version 5.6
               **  Last date of change:   3 Apr 1992  **




               The Lund Monte Carlo - JETSET version 7.3
               **  Last date of change:  10 Mar 1992  **

1************* PYINIT: initialization of PYTHIA routines ********

 =================================================================
 I                                                               I
 I          PYTHIA will be initialized for a p on p collider     I
 I              at    200.000 GeV center-of-mass energy          I
 I                                                               I
 =================================================================

 ** PYMAXI: summary of differential cross-section maximum search *

      ==========================================================
      I                                      I                 I
      I  ISUB  Subprocess name               I  Maximum value  I
      I                                      I                 I
      ==========================================================
      I                                      I                 I
      I  191   f + f~ -> f + f~              I    6.5589E-05   I
      I                                      I                 I
      ==========================================================

 *************** PYINIT: initialization completed ****************



                            Event listing (summary)

    I  particle/jet  KF orig   p_x     p_y     p_z      E      m

    1  !p+!       2212^   0   0.000   0.000  99.996  100.000  0.938
    2  !p+!       2212v   0   0.000   0.000 -99.995  100.000  0.938
 ==================================================================
    3  !u!           2^   1  -0.414  -0.069  25.831   25.835  0.000
    4  !u~!         -2v   2  -0.207  -0.420   0.321    0.568  0.000
    5  !u!           2^   3  -0.414  -0.069  25.831   25.835  0.000
    6  !u~!         -2v   4  -0.207  -0.420   0.321    0.568  0.000
    7  !u!           2    0  -1.025   0.598  23.197   23.227  0.006
    8  !u~!         -2    0   0.404  -1.087   2.956    3.175  0.006
 ==================================================================
    .....


 Warning: maximum violated by  1.017E+00 in event      10
 ISUB = 191; Point of violation:
 tau=4.112E-04, y*=-1.309E+00, cthe=-0.6715451, tau'=0.000E+00,
 phi = 1.299
 XSEC(191,1) increased to  6.673E-05
 Warning: maximum violated by  1.011E+00 in event      82
 ISUB = 191; Point of violation:
 tau =3.328E-04, y*=-8.523E-01, cthe=0.8431695, tau'=0.000E+00,
 phi = 1.961
 XSEC(191,1) increased to  6.745E-05
 Warning: maximum violated by  1.039E+00 in event     566
 ISUB = 191; Point of violation:
 tau =3.455E-04, y*=-9.051E-01, cthe=0.8579352, tau'=0.000E+00,
 phi = 1.735
 XSEC(191,1) increased to  7.010E-05

1*** PYSTAT:  Statistics on Number of Events and Cross-sections ***

 ==================================================================
 I                               I                    I           I
 I            Subprocess         I   Number of points I   Sigma   I
 I                               I                    I           I
 I-------------------------------I--------------------I   (mb)    I
 I                               I                    I           I
 I N:o Type                      I Generated    Tried I           I
 I                               I                    I           I
 ==================================================================
 I                               I                    I           I
 I   0 All included subprocesses I      1000     4831 I 1.481E-05 I
 I 191 f + f~ -> f + f~          I      1000     4831 I 1.481E-05 I
 I                               I                    I           I
 ==================================================================

 *** Fraction of events that fail fragmentation cuts =  0.00000 ***

\end{verbatim}

{\Large\bf Acknowledgements}
\newline
\vspace{0.2cm}\newline
The authors are very grateful to T.~Sj\"ostrand for the excellent
support for the {\sc Pythia/Jetset} code. Furthermore we thank
T.~Gehrmann for providing us an evolution program.
This work was supported by BMBF and DFG (G.~Hess program).
A.S. thanks also the MPI for nuclear physics, Heidelberg for its support.


\begin{thebibliography}{9}
\bibitem{sjo}
T. Sj\"ostrand and M. Bengtsson, Comp. Phys. Commun. 43 (1987) 367.
\newline
H.-U. Bengtsson and T. Sj\"ostrand, Comp. Phys. Commun. 46 (1987) 43.

\bibitem{sjo2} T. Sj\"ostrand, Preprint, hep-ph/9508391.

\bibitem{jaf} R.L. Jaffe, lecture notes, hep-ph/9602236, and
references therein.

\bibitem{bnl} RHIC Spin Collaboration, Proposal on spin physics
using the RHIC polarized collider, (1992).

\bibitem{guelle} S. G\"ullenstern et. al., Comp. Phys. Commun. 87
(1995) 416-431.

\bibitem{ral} J. Ralston and D.E. Soper, Nucl. Phys. B152 (1979) 109.

\bibitem{bartat} J. Bartelski and S. Tatur, Acta. Phys. Polon B27 (1996)
911-920.

\bibitem{gehsti} T. Gehrmann and W.J. Stirling, Phys. Rev. D53 (1996)
6100-6109.
\end{thebibliography}
\end{document}